\documentclass[a4paper,11pt]{article}
\usepackage{amsmath,amssymb}
\usepackage[english]{babel}
\usepackage[pdfpagemode=UseOutlines,
 bookmarks=true,bookmarksopen=true,bookmarksnumbered=true,
 pdffitwindow=true,pdfpagelayout=SinglePage,pdfhighlight=/P,
 colorlinks=true,linkcolor=blue,citecolor=blue,urlcolor=blue,
 unicode=true]{hyperref}

\def\iu{\mathop{\text{\rmfamily\bfseries i}}\nolimits}
\def\const{\mathop{\rm const}\nolimits}
\def\diag{\mathop{\rm diag}\nolimits}

\title{Quantum tops as examples of commuting differential operators}
\author{V.E. Adler, V.G. Marikhin, A.B. Shabat}
\date{26 September 2011}

\begin{document}\maketitle

\noindent\hrulefill
\par\noindent {\bf Abstract.}\quad
We study the quantum analogs of tops on Lie algebras $so(4)$ and $e(3)$
represented by differential operators.
\par\noindent\hrulefill
\smallskip

\section{Introduction}

The commutative rings of differential operators play an important role in
mathematical physics. The problem of their description, in the case of one
independent variable, was set up and solved in the works by Schur \cite{Schur}
and Burchnall--Chaundy \cite{Burchnal_Chaundy_1928}. Later on, these results
have found numerous applications in the theory of integrable equations of the
Korteweg--de Vries type; we mention only the description of finite-gap
operators obtained by Novikov, Krichever and others
\cite{Novikov_1974,Krichever_1978}, and the algorithm of checking necessary
integrability conditions developed by Shabat and others
\cite{Mikhailov_Shabat_Yamilov_1987}.

In the case of two independent variable, the generalizations in different
directions are possible. In general, the problem on a pair of commuting
operators $[\hat H,\hat K]=0$ is not very well posed and the question about
additional assumptions arises. Apart from the conditions suggested by the logic
of the problem itself (for instance, commutativity of a triple of operators),
it is also natural to select the classes of operators which are important in
applications. Here we consider one such class related to the quantum tops on
$so(4)$ and $e(3)$. The corresponding operator $\hat H$ is of a very special
form
\[
 \hat H=a(x)D^2_x+2f(x,y)D_xD_y+b(y)D^2_y+\dots
\]
where $a,b$ are fourth degree polynomials, $f(x,y)$ is a biquadratic polynomial
and dots denote the lower order terms, also with special coefficients. This
form appears if we consider a Hamiltonian which is quadratic with respect to
the generators of Lie algebra, for instance
\[
 \hat H=(\hat U,A\hat U)+(\hat V,B\hat V)+2(\hat U,F\hat V)
  +(\vec c,\hat U)+(\vec d,\hat V)
\]
in the case of $so(4)$, and use a differential representation such that
generators $U,V$ are replaced with differential operators of first order with
quadratic coefficients (see equations (\ref{hatUx}), (\ref{so4.hatHD}) below).
The order of operator $\hat K$ does not exceed 4 in the known examples and,
respectively, the degree of its coefficients does not exceed 8. The original
classical version of the problem deals with Hamiltonians polynomial in momenta
which are in involution with respect to Poisson--Darboux bracket, instead of
commuting differential operators. The conditions of ``commutativity of the
principal part'' \cite{Gabiev_Shabat_2011}, that is, cancellation of the
leading terms in the commutator, coincide for the classical and quantum cases.
In more details, this relation is illustrated by the diagram
\[
\begin{array}{ccc}
 \{H,K\}=0 & \xrightarrow{\quad\text{quantization on the Lie algebra}\quad} &
  [\hat H,\hat K]=0\\[1ex]
 R^*~~\Big\downarrow~~ && ~~\Big\downarrow~~\hat R^* \\[1ex]
 \{H_D,K_D\}=0 & \xrightarrow{\quad\text{quantization in the Darboux coordinates}\quad} &
  [\hat H_D,\hat K_D]=0
\end{array}
\]
Here $R$ denotes a representation of the Lie--Poisson bracket in the Darboux
coordinates, and $H_D$ is the image of the Hamiltonian $H$ under the pullback
map, that is $H_D=R^*(H)=H\circ R$. Analogously, $\hat R$ denotes a
representation of the Lie algebra of differential operators and $\hat H_D=\hat
R^*(\hat H)=\hat H\circ\hat R$. The horizontal arrows correspond to the passage
from the Lie--Poisson bracket to the commutator on the Lie algebra and from the
Poisson--Darboux bracket to the Heisenberg algebra
\[
 \{p_i,x_i\}=\delta_{ij} \quad\to\quad [D_{x_i},x_j]=\delta_{ij}.
\]
This passage, on both levels, is performed by choosing a suitable ordering of
the monomials of Hamiltonians $H,K$, which is equivalent to adding the lower
order terms (``quantum corrections'').

Let us remind that the description of the classical tops admitting an
additional first integral is still an open problem. In the $so(4)$ case, the
necessary condition were derived by Veselov \cite{Veselov_1983} (see
\cite{Reyman_Semenov_1986} for a more precise result) which define a certain
varieties in the parameter space. However, the number of parameters in the
known examples is less at least by 1 than the dimension of these varieties.
Some classification results are presented in paper \cite{Sokolov_2004}. Is it
possible to make progress in this problem, or, more precisely, in its quantum
version, by use of the language of differential operators? As a first step, in
this paper we construct the differential operators which correspond to several
most important examples. In particular, these examples suggest that probably
the simplest necessary condition of integrability is that the discriminant of
the leading part of $\hat H$ is factorable:
\begin{equation}\label{dis}
 f(x,y)^2-a(x)b(y)=w(x,y)\tilde w(x,y)
\end{equation}
where $w,\tilde w$ are, as well as $f$, biquadratic polynomials. Notice that
this relation is important in the problem of separation of variables
\cite{Marikhin_Sokolov_2005}. It is not clear now, how to prove this property
in the general setting, using only the assumption that there exists a commuting
operator $\hat K$ of arbitrary order. From the computational point of view, the
advantages of the differential representation is that it takes care of such
issues as ordering in the operator algebra and Casimir operators. Notice that
representation $\hat R$ is defined on a fixed level set of Casimir operators
and their values occur in $\hat H_D,\hat K_D$ as independent parameters (their
number is equal to the rank of Lie algebra, that is 2 in $so(4)$ and $e(3)$
cases). Splitting of the commutation relation with respect to these parameters
makes its analysis more easy.

Contents of the paper are the following. The $so(4)$ case considered in section
\ref{s:so4} includes the tops of: Schottky--Manakov \cite{Schottky_1891,
Manakov_1976}, Stekloff \cite{Stekloff_1909}, Adler--van Moerbeke
\cite{Adler_Moerbeke_1986, Reyman_Semenov_1986} and Sokolov
\cite{Sokolov_2002}. The quantization of the Schottky--Manakov top was studied
in paper \cite{Komarov_Kuznetsov_1991}. The cases of Adler--van Moerbeke and
Sokolov are more complicated (the operator $\hat K$ is of fourth order) and, up
to our knowledge, their quantization is obtained here for the first time.

In section \ref{s:e3}, we consider tops on $e(3)$: the Clebsch top
\cite{Clebsch_1870}, the Kowalevskaya top \cite{Kowalevski_1889} (quantization
is due to Laporte \cite{Laporte_1933}), the Kowalevskaya gyrostat and the
Goryachev--Chaplygin case. The quantization of models on $e(3)$ is rather well
studied by Komarov and others \cite{Komarov_1981, Komarov_1982,
Komarov_Kuznetsov_1987, Komarov_2001, Ramani_Grammaticos_Dorizzi_1984}.

Certainly, the above examples do not exhaust all known cases. An account of the
known results and a detailed bibliography can be found in book
\cite{Borisov_Mamaev_2005}. Other directions in the problem on commuting
differential operators were developed by many authors, we mention only papers
\cite{Hietarinta_1998, Olshanetsky_Perelomov_1983}.

\paragraph{Notations.} We denote $x=x_1$, $y=x_2$ for short. The definitions of
Poisson brackets and commutators do no include vanishing brackets. The
subscripts $i,j,k$ always denote an arbitrary permutation of $1,2,3$. The
imaginary unit is denoted as $\iu$. The notations
\[
 \vec a=(a_1,a_2,a_3),\quad (\vec a,\vec b)=a_1b_1+a_2b_2+a_3b_3
\]
are used. The following polynomials are often used in writing a compact form of
Hamiltonians in Darboux coordinates:
\begin{equation}\label{WR}
\begin{gathered}[b]
 W(a,b,c;x,y)=a(x^2-1)(y^2-1)-b(x^2+1)(y^2+1)+4cxy,\\
 R(a,b,c;x)=W(a,b,c,x,x)=(a-b)(x^4+1)+2(2c-a-b)x^2.
\end{gathered}
\end{equation}
Notice that operator $W(a,b,c;x,y)D_xD_y$ is form invariant under the inversion
$x\to1/x$, $y\to1/y$.

\section{Tops on \texorpdfstring{$so(4)\simeq so(3)\oplus so(3)$}{so(4)=so(3)+so(3)}}
\label{s:so4}

\subsection{Darboux coordinates and operator representation}

The Lie--Poisson bracket on $so(3)$ and its Casimir function are defined by
equations
\begin{equation}\label{UU}
 \{U_i,U_j\}=\iu\varepsilon_{ijk}U_k,\quad (U,U)=s_1^2
\end{equation}
where $U$ denotes the vector $(U_1,U_2,U_3)$. A representation in Darboux
coordinates is given by relations
\begin{equation}\label{Ux}
 \begin{gathered}[b]
 U_1=-\frac{1}{2}(x^2-1)p_1+s_1x,\quad
 U_2=-\frac{\iu}{2}(x^2+1)p_1+\iu s_1x,\\
 U_3=-xp_1+s_1.
\end{gathered}
\end{equation}
In order to quantize, we pass to the operator algebra
\begin{equation}\label{hatUU}
 [\hat U_i,\hat U_j]=\iu\varepsilon_{ijk}\hat U_k,\quad
 (\hat U,\hat U)=j_1(j_1+1)
\end{equation}
which is represented by differential operators as follows:
\begin{equation}\label{hatUx}
 \begin{gathered}[b]
 \hat U_1=-\frac{1}{2}(x^2-1)D_x+j_1x,\quad
 \hat U_2=-\frac{\iu}{2}(x^2+1)D_x+\iu j_1x,\\
 \hat U_3= -xD_x+j_1.
\end{gathered}
\end{equation}
The Lie algebra $so(4)\simeq so(3)\oplus so(3)$ is represented as the direct
sum with a double under the following renaming:
\[
 U\to V,\quad x\to y,\quad p_1\to p_2,\quad s_1\to s_2,\quad
 \hat U\to \hat V,\quad D_x\to D_y,\quad j_1\to j_2.
\]
The tops on $so(4)$ are defined by Hamiltonians of the general form
\begin{equation}\label{so4.hatH}
 \hat H=(U,AU)+(V,BV)+(U,FV)+(\vec c,U)+(\vec d,V).
\end{equation}
The matrices $A$ and $B$ can be chosen diagonal without loss of generality. The
use of representation (\ref{hatUx}) yields a differential operator of the
special form
\begin{equation}\label{so4.hatHD}
\begin{aligned}[b]
 \hat H_D &= a(x)D^2_x+f(x,y)D_xD_y+b(y)D^2_y\\
  &\quad -\Bigl(\frac{2j_1-1}{2}a'(x)+2c(x)+j_2f_y(x,y)\Bigr)D_x\\
  &\quad -\Bigl(\frac{2j_2-1}{2}b'(y)+2d(y)+j_1f_x(x,y)\Bigr)D_y\\
  &\quad +\frac{j_1(2j_1-1)}{6}a''(x)+\frac{j_2(2j_2-1)}{6}b''(y)\\
  &\quad +2j_1c'(x)+2j_2d'(y)+j_1j_2f_{xy}(x,y)+\kappa
\end{aligned}
\end{equation}
where $\kappa$ is an arbitrary constant and
\begin{gather*}
 a(x)=a_0(x^4+1)+a_2x^2,\quad b(y)=b_0(y^4+1)+b_2y^2,\\
 c(x)=c_2x^2+c_1x+c_0,\quad d(y)=d_2y^2+d_1y+d_0,\\
 f(x,y)=f_{22}x^2y^2+\dots+f_{00}.
\end{gather*}
The correspondence between (\ref{so4.hatHD}) and (\ref{so4.hatH}) is given by
relations
\begin{gather*}
 A=\diag(2a_0,-2a_0,a_2)+\alpha I,\quad
 B=\diag(2b_0,-2b_0,b_2)+\beta I,\\
 \vec c=2(c_2-c_0,-\iu(c_2+c_0),c_1),\quad
 \vec d=2(d_2-d_0,-\iu(d_2+d_0),d_1),\\
 F=\begin{pmatrix}
  f_{00}-f_{02}-f_{20}+f_{22} & \iu(f_{00}+f_{02}-f_{20}-f_{22} &
     f_{21}-f_{01} \\
  \iu(f_{00}-f_{02}+f_{20}-f_{22}) & -f_{00}-f_{02}-f_{20}-f_{22} &
    -\iu(f_{21}+f_{01}) \\
  f_{12}-f_{10} & -\iu(f_{12}+f_{10}) & f_{11}
 \end{pmatrix},\\
 3\kappa=(3\alpha+a_2)j_1(j_1+1)+(3\beta+b_2)j_2(j_2+1)
\end{gather*}
with arbitrary constants $\alpha,\beta$.

The following remark is useful for computing spectra. It is about the relation
between differential representation (\ref{hatUx}) and the matrix representation
\cite[p. 88]{LL}
\begin{equation}\label{M}
\begin{aligned}[b]
 &\hat U_1\pm\iu \hat U_2=\hat U_\pm,\quad \hat U_3|m,j\rangle =m|m,j\rangle,\\
 &\hat U_+|m,j\rangle =\sqrt{(j-m)(j+m+1)}|m+1,j\rangle,\\
 &\hat U_-|m,j\rangle =\sqrt{(j+m)(j-m+1)}|m-1,j\rangle.
\end{aligned}
\end{equation}
In the physical case we consider an orbit $j=\const$ such that $j$ and $m$ are
simultaneously integer or half-integer and $m$ takes the values from $-j$ to
$j$, which corresponds to the matrices of the finite size $(2j+1)\times(2j+1)$.
Formally, one can waive these restrictions and consider the infinite
tridiagonal matrices which are interpreted as the difference operators
\[
 a(m)T_m+b(m)+c(m)T^{-1}_m
\]
where $T_m:m\mapsto m+1$ is the shift operator (the matrix with units on the
subdiagonal). A conjugation $\hat A\to f^{-1}\hat Af$ allows to get rid of
square roots, and the representation of $so(3)$ by shift operators appears:
\begin{equation}\label{T}
\begin{aligned}[b]
 \hat U_1 &= -\frac{1}{4}(m-j)(m+j+1)T_m +T^{-1}_m,\\
 \hat U_2 &= -\iu\Bigl(\frac{1}{4}(m-j)(m+j+1)T_m +T^{-1}_m\Bigr),\\
 \hat U_3 &= m.
\end{aligned}
\end{equation}
Another difference representation can be obtained directly from (\ref{hatUx})
via the change (``Fourier transformation'')
\[
 D_x\to mT_m,\quad x\to T^{-1}_m
\]
which preserves the Heisenberg algebra:
\[
 [D_x,x]=1, \quad [mT_m,T^{-1}_m]=1,
\]
and yields the operators
\begin{equation}\label{T1}
\begin{aligned}[b]
 \hat U_1 &= \frac{m}{2}T_m +\left(j+1-\frac{m}{2}\right)T^{-1}_m,\\
 \hat U_2 &= -\iu\frac{m}{2}T_m +\iu\left(j+1-\frac{m}{2}\right)T^{-1}_m,\\
 \hat U_3 &= j-m+1.
\end{aligned}
\end{equation}
One can check that representations (\ref{T}) and (\ref{T1}) are equivalent up
to a certain linear transformation from $SO(3)$ and the conjugation by operator
$g(m)T^{-j-1}$ where function $g$ satisfies relation $g(m+2)/g(m+1)=2j-m$.

\subsection{Schottky--Manakov top}\label{s:SM}

The commuting operators are of the form
\begin{align}
\nonumber
 \hat H&= -\alpha_1^2\hat U_1^2-\alpha_2^2\hat U_2^2-\alpha_3^2\hat U_3^2
     -\alpha_1^2\hat V_1^2-\alpha_2^2\hat V_2^2-\alpha_3^2\hat V_3^2\\
\label{SM.hatH}
  & \qquad +2\alpha_2\alpha_3\hat U_1\hat V_1
           +2\alpha_3\alpha_1\hat U_2\hat V_2
           +2\alpha_1\alpha_2\hat U_3\hat V_3,\\
\label{SM.hatK}
 \hat K&= \alpha_1\hat U_1\hat V_1+\alpha_2\hat U_2\hat V_2+\alpha_3\hat U_3\hat V_3.
\end{align}
These can be obtained from the classical Hamiltonians in involution by the
simple exchange $U\to\hat U$, $V\to\hat V$, because the problem of normal
ordering does not appear in this example (each monomial contains commuting
variables only). In contrast to some other examples (Stekloff, Kowalevskaya
tops), no linear terms can be added to the Hamiltonians, both in classical and
quantum cases, as one can prove by a direct computation with indeterminate
coefficients. Applying map (\ref{hatUx}) yields the following pair of commuting
differential operators:
\begin{align}
\label{SM.hatHd}
&\begin{aligned}[b]
 &4\hat H_D= r(x)D^2_x+2zD_xD_y+r(y)D^2_y\\
 &\quad -\left(\frac{2j_1-1}{2}r'(x)+2j_2z_y\right)D_x
          -\left(\frac{2j_2-1}{2}r'(y)+2j_1z_x\right)D_y\\
 &\quad +2j_1j_2z_{xy}
        +\frac{j_1(2j_1-1)}{6}r''(x)+\frac{j_2(2j_2-1)}{6}r''(y)\\
 &\quad -\frac{4}{3}(j_1(j_1+1)+j_2(j_2+1))(\alpha^2_1+\alpha^2_2+\alpha^2_3),
\end{aligned}\\
\label{SM.hatKd}
 &4\hat K_D=wD_xD_y-j_2w_yD_x-j_1w_xD_y+j_1j_2w_{xy}
\end{align}
where, using the notations (\ref{WR}),
\begin{align*}
 & r(x)=-R(\alpha_1^2,\alpha_2^2,\alpha_3^2;x),\\
 & z=z(x,y)=W(\alpha_2\alpha_3,\alpha_3\alpha_1,\alpha_1\alpha_2;x,y),\\
 & w=w(x,y)=W(\alpha_1,\alpha_2,\alpha_3;x,y).
\end{align*}
The constant in the last line of (\ref{SM.hatHd}) ($\kappa$ in equation
(\ref{so4.hatHD})) has no effect on the commutativity and determines the energy
shift of the ground state of the system.

Alternatively, operators (\ref{SM.hatHd}), (\ref{SM.hatKd}) can be obtained by
another route along the diagram from Introduction. First, one should pass to
the Darboux coordinates in the classical Hamiltonians $H,K$, accordingly to
formulae (\ref{Ux}). This yields
\begin{align}
\label{SM.Hd}
&\begin{aligned}[b]
  &4H_D= r(x)p^2_1+2zp_1p_2+r(y)p^2_2\\
  &\quad -(s_1r'(x)+2s_2z_y)p_1-(s_2r'(y)+2s_1z_x)p_2\\
  &\quad +2s_1s_2z_{xy}
         +\frac{s^2_1}{3}r''(x)+\frac{s^2_2}{3}r''(y)
         -\frac{4}{3}(s^2_1+s^2_2)(\alpha^2_1+\alpha^2_2+\alpha^2_3),
\end{aligned}\\
\label{SM.Kd}
 & 4K_D=wp_1p_2-s_2w_yp_1-s_1w_xp_2+s_1s_2w_{xy}
\end{align}
where $r,z,w$ are the same as above. Next, $p_i$ are replaced by $D_{x_i}$ in
quadratic terms and the lower order terms are taken with indeterminate
coefficients (preserving the degree with respect to $x,y$). So, one searches
for the commuting differential operators of the form
\begin{gather*}
 \hat H_D= r(x)D^2_x+2z(x,y)D_xD_y+r(y)D^2_y
  +a(\overset{3}{x},\overset{1}{y})D_x
  +b(\overset{1}{x},\overset{3}{y})D_y
  +c(\overset{2}{x},\overset{2}{y}),\\
 \hat K_D=w(x,y)D_xD_y
  +f(\overset{2}{x},\overset{1}{y})D_x
  +g(\overset{1}{x},\overset{2}{y})D_y
  +h(\overset{1}{x},\overset{1}{y})
\end{gather*}
where $r,z,w$ are given, $a,b,c,f,g,h$ are polynomials with indeterminate
coefficients, and numbers above arguments show the corresponding degrees. The
resulting system of equations is rather bulky, but its solution
(\ref{SM.hatHd}), (\ref{SM.hatKd}) can be easily recovered by any system of
computer algebra. Thus, both methods of quantization turn out to be equivalent,
as one should expect. In the situation, when $\hat H$ is given and $\hat K$ is
to be found, each method has its own strong and weak points from the point of
view of computation complexity. The use of the Lie algebra generators gives a
simpler equations for the coefficients, but its implementation is more
difficult because it deals with noncommutative variables. The language of
differential representation is more flexible and universal, but the resulting
system for the coefficients of the operators is slightly more cumbersome.

\subsection{Stekloff top}\label{s:St}

The classical Hamiltonians are of the form ($\alpha=\alpha_1\alpha_2\alpha_3$)
\begin{align*}
 H&= -\alpha^2\left(
       \frac{1}{\alpha_1^2}U_1^2
      +\frac{1}{\alpha_2^2}U_2^2
      +\frac{1}{\alpha_3^2}U_3^2\right)
     +2\alpha(\alpha_1U_1V_1
             +\alpha_2U_2V_2
             +\alpha_3U_3V_3),\\
 K&= 2\alpha\left(
       \frac{1}{\alpha_1}U_1V_1
      +\frac{1}{\alpha_2}U_2V_2
      +\frac{1}{\alpha_3}U_3V_3\right)
     -\alpha_1^2V_1^2-\alpha_2^2V_2^2-\alpha_3^2V_3^2.
\end{align*}
Like in the previous example, the problem of ordering does not appear and the
commuting quantum Hamiltonians are obtained just by placing hats over $U,V$. It
is easy to check that the following linear terms can be added to the
Hamiltonians, both in the classical and quantum cases:
\begin{gather*}
 H\to H-\beta_1\alpha_2\alpha_3U_1
       -\beta_2\alpha_1\alpha_3U_2
       -\beta_3\alpha_1\alpha_2U_3,\\
 K\to K+\beta_1V_1+\beta_2V_2+\beta_3V_3,
\end{gather*}
with arbitrary $\beta_i$. However, we assume $\beta_i=0$ for the sake of
simplicity.

The passage to the differential operators yields
\begin{align*}
&\begin{aligned}[b]
 &4\hat H_D= r_1D^2_x+2zD_xD_y-\left(\frac{2j_1-1}{2}r'_1+2j_2z_y\right)D_x-2j_1z_xD_y\\
 &\quad +2j_1j_2z_{xy}+\frac{j_1(2j_1-1)}{6}r''_1
 -\frac{4}{3}j_1(j_1+1)
  (\alpha^2_1\alpha^2_2+\alpha^2_2\alpha^2_3+\alpha^2_3\alpha^2_1),
\end{aligned}\\
&\begin{aligned}[b]
 &4\hat K_D= 2wD_xD_y+r_2D^2_y-2j_2w_yD_x-\left(\frac{2j_2-1}{2}r'_2+2j_1w_x\right)D_y\\
 &\quad +2j_1j_2w_{xy}+\frac{j_2(2j_2-1)}{6}r''_2
   -\frac{4}{3}j_2(j_2+1)(\alpha^2_1+\alpha^2_2+\alpha^2_3)
\end{aligned}
\end{align*}
where
\begin{align*}
 r_1&=r_1(x)=-R(\alpha^2_2\alpha^2_3,\alpha^2_3\alpha^2_1,\alpha^2_1\alpha^2_2;x),\\
 r_2&=r_2(y)=-R(\alpha^2_1,\alpha^2_2,\alpha^2_3;y),\\
   z&=z(x,y)=\alpha_1\alpha_2\alpha_3W(\alpha_1,\alpha_2,\alpha_3;x,y),\\
   w&=w(x,y)=W(\alpha_2\alpha_3,\alpha_3\alpha_1,\alpha_1\alpha_2;x,y).
\end{align*}
The same operators are obtained via the quantization in the Darboux variables,
starting from the Hamiltonians
\begin{align*}
&\begin{aligned}[b]
 4H_D&= r_1p^2_1+2zp_1p_2-(s_1r'_1+2s_2z_y)p_1-2s_1z_xp_2\\
 &\quad +2s_1s_2z_{xy}+\frac{s^2_1}{3}r''_1
  -\frac{4}{3}s^2_1(\alpha^2_1\alpha^2_2+\alpha^2_2\alpha^2_3+\alpha^2_3\alpha^2_1),
\end{aligned}\\
&\begin{aligned}[b]
 4K_D&= 2wp_1p_2+r_2p^2_2-2s_2w_yp_1-(s_2r'_2+2s_1w_x)p_2\\
 &\quad +2s_1s_2w_{xy}+\frac{s^2_2}{3}r''_2
  -\frac{4}{3}s^2_2(\alpha^2_1+\alpha^2_2+\alpha^2_3).
\end{aligned}
\end{align*}

\subsection{M.\,Adler--van Moerbeke top}\label{s:AvM}

The classical Hamiltonians are of the following form (the parameters are
related by the constraint $\lambda_1+\lambda_2+\lambda_3=0$):
\begin{align}
\nonumber
 H&= \sum_{i=1}^3\Bigl(-9\lambda_j^2\lambda_k^2U_i^2
      +6\lambda_j\lambda_k(\lambda_j-\lambda_i)(\lambda_k-\lambda_i)U_iV_i\\
\label{AvM.H}
  &\qquad +\lambda_j\lambda_k(4\lambda_i^2-\lambda_j\lambda_k)V_i^2\Bigr),\\
\nonumber
 K&= 3\sum_{i,j}\lambda_j(\lambda_i-\lambda_j)U_iV_iV_j^2
   +\sum_i(\lambda_i-\lambda_j)(\lambda_i-\lambda_k)U_iV_i^3\\
\label{AvM.K}
 &\qquad -9\sum_lU_l^2\sum_i\lambda_j\lambda_kU_iV_i
 +\frac{3}{2}\sum_lU_l^2\sum_i(3\lambda_i^2-\lambda_j^2-\lambda_k^2)V_i^2.
\end{align}
The quantum Hamiltonian $\hat H$ is obtained from $H$ by fixing hats. However,
$\hat K$ contains monomials with noncommutative generators and the problem of
ordering arises. The result of direct computation with indeterminate
coefficients is the following expression for $\hat K$:
\begin{equation}\label{AvM.hatK}
\begin{aligned}[b]
 \hat K&= \sum_{i,j}\lambda_j(\lambda_i-\lambda_j)
    \bigl(\hat{U}_i\hat{V}_i\hat{V}_j^2
         +\hat{U}_i\hat{V}_j\hat{V}_i\hat{V}_j
         +\hat{U}_i\hat{V}_j^2\hat{V}_i\bigr)\\
 &\quad
  +\sum_i(\lambda_i-\lambda_j)(\lambda_i-\lambda_k)\hat{U}_i\hat{V}_i^3
  -9\Bigl(\sum_l\hat{U}_l^2+\frac{1}{3}\Bigr)\sum_i\lambda_j\lambda_k\hat{U}_i\hat{V}_i\\
 &\quad
 +\frac{3}{2}\sum_l\hat{U}_l^2\sum_i(3\lambda_i^2-\lambda_j^2-\lambda_k^2)\hat{V}_i^2.
\end{aligned}
\end{equation}
Thus, the quantum Hamiltonian differs from the classical one by ordering in the
first sum and a quantum correction in the third sum (recall, that $\sum_l\hat
U_l^2$ is the Casimir function).

The following expression for $\hat H_D$ is obtained by passage to differential
operators according to general formula (\ref{so4.hatHD}):
\begin{equation}\label{AvM.hatHD}
\begin{aligned}[b]
 4\hat H_D &= r_1(x)D^2_x+2zD_xD_y+r_2(y)D^2_y\\
  &\quad -\Bigl(\frac{2j_1-1}{2}r'_1(x)+2j_2z_y\Bigr)D_x
         -\Bigl(\frac{2j_2-1}{2}r'_2(y)+2j_1z_x\Bigr)D_y\\
  &\quad
   +\frac{j_1(2j_1-1)}{6}r''_1(x)+\frac{j_2(2j_2-1)}{6}r''_2(y) +2j_1j_2z_{xy}\\
  &\quad
  -\frac{4}{3}(9j_1(j_1+1)+j_2(j_2+1))(\lambda^2_1+\lambda_1\lambda_2+\lambda^2_2)^2
\end{aligned}
\end{equation}
where
\begin{align*}
 & r_1(x)=-9R(\lambda^2_2\lambda^2_3,\lambda^2_3\lambda^2_1,\lambda^2_1\lambda^2_2,x),\\
 & r_2(x)=\frac{1}{9}r_1(x)+4\lambda_1\lambda_2\lambda_3R(\lambda_1,\lambda_2,\lambda_3,x),\\
 & z=z(x,y)=3W(\mu_1,\mu_2,\mu_3,x,y),\quad
   \mu_i=(\lambda^2_i-\lambda^2_j)(\lambda^2_i-\lambda^2_k).
\end{align*}
The structure of operator $\hat K_D$ is rather simple:
\begin{align*}
 4\hat K_D&= g^2D_xD^3_y+c_1gg_yD_xD^2_y+c_2gg_xD^3_y \\
  &\quad +(c_3gg_{yy}+c_4g^2_y)D_xD_y+(c_5g_yg_x+c_6gg_{xy})D^2_y \\
  &\quad +(c_7gg_{yyy}+c_8g_yg_{yy})D_x+(c_9g_{yy}g_x+c_{10}g_yg_{xy}+c_{11}gg_{xyy})D_y \\
  &\quad +c_{12}g_{yyy}g_x+c_{13}g_{yy}g_{xy}+c_{14}g_yg_{xyy}+c_{15}gg_{xyyy},
\end{align*}
but the coefficients are cumbersome. Here $g$ denotes the polynomial
\[
 g=(\lambda_1-\lambda_2)(xy^3+1)+3\lambda_3(x+y)y
\]
related to the coefficients of $\hat H_D$ by relation
\[
 z^2-r_1(x)r_2(y)=36\lambda_1\lambda_2\lambda_3
  (\lambda_1-\lambda_2)(\lambda_2-\lambda_3)(\lambda_3-\lambda_1)g\tilde g,
\]
where $\tilde g=(\lambda_1-\lambda_2)x(x+y)-\lambda_3(x^3y+1)$. The
coefficients $c_i$ are, in turn, polynomials in the parameters $j_1,j_2$:
\begin{align*}
 & c_1=-2(j_2-1),\quad c_2=-2j_1,\\
 & c_3=\frac{1}{2}(3j_1(j_1+1)+j_2(j_2-3)+2),\quad c_4=-j_1(j_1+1)+j_2(j_2-1), \\
 & c_5= 2j_1(j_1+j_2),\quad c_6=-2j_1(j_1-j_2+2), \\
 & c_7= -\frac{j_2}{6}(9j_1(j_1+1)-j_2(j_2+3)+4),\quad
   c_8= \frac{j_2}{2}(j_1(j_1+1)-j_2(j_2-1)),\\
 & c_9= -\frac{j_1}{2}(3j^2_1+4j_1j_2+j^2_2+j_1+j_2),\quad
   c_{10}= 2j_1(j_1(j_1+1)-j_2(j_2-1)),\\
 & c_{11}=-\frac{j_1}{2}(3j^2_1-4j_1j_2+j^2_2+5j_1-7j_2+4), \\
 & c_{12}=\frac{j_1j_2}{6}(9j^2_1+3j_1j_2-j^2_2+6j_1+1),\\
 & c_{13}=\frac{j_1j_2}{6}(-3j^2_1+5j_1j_2+3j^2_2-4j_1+2j_2-1),\\
 & c_{14}=\frac{j_1j_2}{6}(-3j^2_1-5j_1j_2+3j^2_2-2j_1-8j_2+1),\\
 & c_{15}=\frac{j_1j_2}{6}(9j^2_1-3j_1j_2-j^2_2+12j_1-6j_2+7).
\end{align*}

\subsection{Sokolov top}\label{s:S}

In this example, it is convenient to use the variables
\[
 m_i=U_i+V_i,\quad n_i=U_i-V_i.
\]
The Sokolov top on $so(4)$ is defined by the Hamiltonian
\begin{equation}\label{S.H}
 H= \frac{1}{2}m_1^2+\frac{1}{2}m_2^2+m_3^2+m_3(\alpha n_1+\beta n_2)
    -\frac{1}{2}(\alpha^2+\beta^2)n_3^2
\end{equation}
and additional fourth order integral is of the form
\begin{equation}\label{S.K}
\begin{gathered}[b]
 K= m_3^2\bigl(2H-m_3^2+(\beta m_1-\alpha m_2)^2+(\alpha n_1+\beta n_2)^2\bigr)\\
 \qquad +2m_3\bigl(\alpha m_1+\beta m_2-(\alpha^2+\beta^2)n_3\bigr)(m_1n_1+m_2n_2).
\end{gathered}
\end{equation}
The quantum Hamiltonian is obtained by symmetrization of noncommutative
monomials:
\begin{multline}\label{S.Hq}
 \hat H= \frac{1}{2}\hat m_1^2+\frac{1}{2}\hat m_2^2+\hat m_3^2
   +[\hat m_3,\alpha\hat n_1+\beta\hat n_2]^+
   -\frac{1}{2}(\alpha^2+\beta^2)\hat n_3^2
\end{multline}
where
\[
 [a,b]^+=\frac12(ab+ba).
\]
The operator $\hat K$ is obtained by symmetrization as well, with some special
weight coefficients. It should be noted that there exist homogeneous
polynomials in the algebra $so(4)$ which vanish identically in virtue of the
commutation relations, so that the form of the operator $\hat K$ is not unique
and one can try to simplify it by adding such polynomials. We write down one of
possible versions. Let us denote
\begin{align*}
 A_{a,b,c}(m,f)&=amf^2m+b[m^2,f^2]^+ +c[f,mfm]^+\\
 &\qquad +(1-a-b-c)fm^2f,\\
 B(m,f,g)&=mfg+gfm,
\end{align*}
then the operator
\begin{equation}\label{S.Kq}
\begin{aligned}[b]
 \hat K=&~
     A_{\frac12,0,\frac34}(\hat m_3,\hat m_1)
    +A_{\frac12,0,\frac34}(\hat m_3,\hat m_2)
    -(\alpha^2+\beta^2)\hat m^2_3\hat n^2_3\\
   &+A_{\frac12,\frac14,1}(\hat m_3,\beta\hat m_1-\alpha\hat m_2)
    +A_{-\frac34,-1,\frac72}(\hat m_3,\hat m_3+\alpha\hat n_1+\beta\hat n_2)\\
   &+B(\hat m_3,\alpha\hat m_1+\beta\hat m_2-(\alpha^2+\beta^2)\hat n_3,
       \hat m_1\hat n_1+\hat m_2\hat n_2),\\
   &-\frac12[\hat m_3\hat n_3,[\hat m_3,\alpha\hat m_1+\beta\hat m_2]]
\end{aligned}
\end{equation}
commutes with $\hat H$ and coincides with $K$ if all variables are commutative.

In this example, it is also convenient to modify slightly our differential
representation of $so(4)$, by changing signs $y\to-y$, $D_y\to-D_y$. This
brings to the operators which are symmetric with respect to $x,y$. Let us
denote (in general, the parameters $\alpha,\beta$ are complex)
\begin{gather*}
 \xi_1=\alpha+\iu\beta,\quad
 \xi_2=-\alpha+\iu\beta,\quad
 r(x)=x(\xi_1x+1)(x+\xi_2),\\
 z=z(x,y)=\xi_1xy(x+y)+(x+y)^2+2(1-\xi_1\xi_2)xy+\xi_2(x+y)
\end{gather*}
then $\hat H$ is written in the form (\ref{so4.hatHD}) as follows:
\begin{align*}
 2\hat H_D &= r(x)D^2_x+zD_xD_y+r(y)D^2_y\\
  &\quad -\Bigl(\frac{2j_1-1}{2}r'(x)+j_2z_y\Bigr)D_x
         -\Bigl(\frac{2j_2-1}{2}r'(y)+j_1z_x\Bigr)D_y\\
  &\quad +\frac{j_1(2j_1-1)}{6}r''(x)+\frac{j_2(2j_2-1)}{6}r''(y)+j_1j_2z_{xy}\\
  &\quad +\frac{1}{3}(j_1(j_1+1)+j_2(j_2+1))(\xi_1\xi_2+4).
\end{align*}
Notice that the polynomial $r$ is of degree 3, because Hamiltonian (\ref{S.H})
takes the form (\ref{so4.hatH}) with nondiagonal matrices $A,B$ if one returns
to the variables $U,V$. The second operator is too bulky and we present
explicitly only the leading terms:
\begin{align*}
 \hat K_D&= w^2(xD_x+yD_y)^2D_xD_y-2j_2x^2ww_yD^3_x-2j_1y^2ww_xD^3_y\\
  &\quad -2xw((j_1-1)xw_x+(j_1+j_2-1)w+(2j_2-1)yw_y)D^2_xD_y\\
  &\quad -2yw((2j_1-1)xw_x+(j_1+j_2-1)w+(j_2-1)yw_y)D^2_xD_y+\dots
\end{align*}
where $w=\xi_1xy+x+y+\xi_2$.

\subsection{The classical limit}

The Planck constant is introduced by simple scaling of the generators, so that
commutation relations (\ref{hatUU}) are replaced with
\[
 [\hat U_i,\hat U_j]=\iu\hbar\varepsilon_{ijk}\hat U_k,\quad
 (\hat U,\hat U)=\hbar^2j_1(j_1+1)
\]
and representation (\ref{hatUx}) is replaced with
\begin{gather*}
 \hat U_1=\hbar\Bigl(-\frac{1}{2}(x^2-1)D_x+j_1x\Bigr),\quad
 \hat U_2=\hbar\Bigl(-\frac{\iu}{2}(x^2+1)D_x+\iu j_1x\Bigr),\\
 \hat U_3=\hbar(-xD_x+j_1).
\end{gather*}
The equations for the variables $V$ are changed analogously. The passage to the
classical limit for any quantum operator $\hat A$ is defined according to the
formula
\[
 A=\lim_{\hbar\to0}
 e^{-\frac{\iu}{\hbar}(p_1x+p_2y)}\left(\hat{A}\,
 e^{\frac{\iu}{\hbar}(p_1x+p_2y)}\right)\big|_{j_i=\frac{s_i}{\hbar}},
\]
in particular the commutator bracket and the Casimir functions for each copy of
$so(3)$ are mapped into the Lie--Poisson bracket (\ref{UU}), and the formulae
for the generators themselves are mapped into the Darboux coordinates
representation (\ref{Ux}).

Applying of this procedure to $\hat H_D$, $\hat K_D$ gives the same expressions
for $H_D$, $K_D$ as the intermediate passage to the Darboux coordinates in the
classical Hamiltonians $H$, $K$. This is guaranteed by the ``correspondence
principle'' which is invariant, that is it does not depend explicitly on the
choice of representation of the algebra $so(4)$. The check of the
correspondence principle is trivial for the Schottky--Manakov and Stekloff
tops, because there are no quantum corrections in these systems. In the
Adler--van Moerbeke case, the correct passage to the classical limit is
achieved by changing one term in the expression for $\hat K$ (\ref{AvM.hatK}):
\[
 \Bigl(\sum_l\hat{U}_l^2+\frac{1}{3}\Bigr)\sum_i\lambda_j\lambda_k\hat{U}_i\hat{V}_i
 \quad\to\quad
 \Bigl(\sum_l\hat{U}_l^2+\frac{\hbar^2}{3}\Bigr)\sum_i\lambda_j\lambda_k\hat{U}_i\hat{V}_i.
\]
After this, all terms become homogeneous with respect to $\hbar$, and the limit
$\hbar\to0$ gives rise to the classical Hamiltonian (\ref{AvM.K}).

It should be noted that the Casimir functions $\hbar^2j_i(j_i+1)$ are of the
quantum nature, because $j_i$ take integer/half-integer values. The passage to
the limit $\hbar\to0$ brings to the classical (finite) quantities $s_i=\hbar
j_i$. On the other hand, if we consider spins then the values $j_i$ are finite
and therefore $s_i\to0$, in accordance with a statement that the spin is a pure
quantum concept.

\section{Tops on $e(3)$}\label{s:e3}

\subsection{Darboux coordinates and operator representations}

The Lie--Poisson bracket on $e(3)$ is of the form
\[
 \{M_i,M_j\}=-\varepsilon_{ijk}M_k,\quad
 \{M_i,\gamma_j\}=-\varepsilon_{ijk}\gamma_k,\quad
 \{\gamma_i,\gamma_j\}=0
\]
and the Casimir functions are
\[
 (M,\gamma)=l,\quad (\gamma,\gamma)=a^2
\]
where $M=(M_1,M_2,M_3)$, $\gamma=(\gamma_1,\gamma_2,\gamma_3)$. We use the
following representation in the Darboux coordinates:
\begin{equation}\label{pq-e3}
\begin{gathered}[b]
\begin{aligned}
 M_1 &= -\frac{\iu}{2}(x^2-1)p_1-\frac{\iu}{2}(y^2-1)p_2+\frac{l}{2a}(x-y),\\
 M_2 &= -\frac{1}{2}(x^2+1)p_1-\frac{1}{2}(y^2+1)p_2-\iu\frac{l}{2a}(x-y),\\
 M_3 &= \iu(xp_1+yp_2),
\end{aligned}\\
 \gamma_1=a\frac{1-xy}{x-y},\quad
 \gamma_2=\iu a\frac{1+xy}{x-y},\quad
 \gamma_3=a\frac{x+y}{x-y}.
\end{gathered}
\end{equation}
A representation with real Darboux coordinates \cite{Marikhin_Sokolov_2010}
should be mentioned as well:
\begin{equation}\label{pq-e3-R}
\begin{gathered}[b]
\begin{aligned}
 M_1&= -p_1q_1q_2+\frac{1}{2}p_2(q_1^2-q_2^2-1)
  +\frac{lq_1(q_1^2+q_2^2+1)}{2a(q_1^2+q_2^2)},\\
 M_2&= p_2q_1q_2+\frac{1}{2}p_1(q_1^2-q_2^2+1)
  +\frac{lq_2(q_1^2+q_2^2+1)}{2a(q_1^2+q_2^2)},\\
 M_3&= p_1q_2-p_2q_1,
\end{aligned}\\
 \gamma_1=\frac{2aq_1}{(q_1^2+q_2^2+1)},\quad
 \gamma_2=\frac{2aq_2}{(q_1^2+q_2^2+1)},\quad
 \gamma_3=\frac{a(q_1^2+q_2^2-1)}{(q_1^2+q_2^2+1)}.
\end{gathered}
\end{equation}

Quantization replaces the Lie--Poisson bracket with the $e(3)$ commutator
\begin{equation}\label{e3_comm}
 [\hat{M}_i,\hat{M}_j]=\iu\varepsilon_{ijk}M_k,\quad
 [\hat{M}_i,\hat{\gamma}_j]=\iu\varepsilon_{ijk}\gamma_k,\quad
 [\hat{\gamma}_i,\hat{\gamma}_j]=0
\end{equation}
and the Casimir operators are
\begin{equation}\label{e3_la}
 (\hat\gamma,\hat M)=l,\quad (\hat\gamma,\hat\gamma)=a^2.
\end{equation}
This operator algebra admits the following representation:
\begin{equation}\label{hat-e3}
\begin{gathered}[b]
\begin{aligned}
 \hat{M}_1 &= \frac{1}{2}(1-x^2)D_x+\frac{1}{2}(1-y^2)D_y+\frac{l}{2a}(x-y),\\
 \hat{M}_2 &= \frac{\iu}{2}(1+x^2)D_x+\frac{\iu}{2}(1+y^2)D_y-\iu\frac{l}{2a}(x-y),\\
 \hat{M}_3 &= xD_x+yD_y,
\end{aligned}\\
 \hat{\gamma}_1= a\frac{1-xy}{x-y},\quad
 \hat{\gamma}_2= \iu a\frac{1+xy}{x-y},\quad
 \hat{\gamma}_3= a\frac{x+y}{x-y}.
\end{gathered}
\end{equation}
Notice that all operators are invariant with respect to the change
\[
 x\leftrightarrow y,\quad a\to -a.
\]

A matrix representation can be obtained by introducing the basis function
$|\psi\rangle=|m,n\rangle=(x+y)^m(x-y)^n$. This choice is motivated by the
denominator of generators $\gamma_i$ in representation (\ref{hat-e3}) and the
symmetry arguments. Easy computation yields ($M_{\pm}=M_1\pm\iu M_2$,
$\gamma_{\pm}=\gamma_1\pm\iu\gamma_2$)
\begin{gather*}
 \hat{M}_3|m,n\rangle = (m+n)|m,n\rangle,\quad
 \hat M_-|m,n\rangle = 2m|m+1,n-1\rangle,\\
 \hat{M}_+|m,n\rangle = -\bigl(n+\frac{m}{2}\bigr)|m+1,n\rangle
 -\frac{m}{2}|m-1,n+2\rangle+\frac{l}{a}|m,n+1\rangle,\\
 \hat\gamma_3|m,n\rangle = a|m+1,n-1\rangle,\quad
 \hat\gamma_-|m,n\rangle = 2a|m,n-1\rangle,\\
 \hat\gamma_+|m,n\rangle = \frac{a}{2}|m,n+1\rangle-\frac{a}{2}|m+2,n-1\rangle.
\end{gather*}

In order to pass to the classical limit, the Planck constant is introduced as
follows:
\[
 [\hat{M}_i,\hat{M}_j]=\iu\hbar\varepsilon_{ijk}M_k,\quad
 [\hat{M}_i,\hat{\gamma}_j]=\iu\hbar\varepsilon_{ijk}\gamma_k,\quad
 [\hat{\gamma}_i,\hat{\gamma}_j]=0,
\]
and the operator representation is replaced by equations
\begin{gather*}
\begin{aligned}
 \hat{M}_1 &= \hbar\Bigl(\frac{1}{2}(1-x^2)D_x
                   +\frac{1}{2}(1-y^2)D_y\Bigr)+\frac{l}{2a}(x-y),\\
 \hat{M}_2 &= \hbar\Bigl(\frac{\iu}{2}(1+x^2)D_x
                   +\frac{\iu}{2}(1+y^2)D_y\Bigr)-\iu\frac{l}{2a}(x-y),\\
 \hat{M}_3 &= \hbar(xD_x+yD_y),
\end{aligned}\\
 \hat{\gamma}_1= a\frac{1-xy}{x-y},\quad
 \hat{\gamma}_2= \iu a\frac{1+xy}{x-y},\quad
 \hat{\gamma}_3= a\frac{x+y}{x-y}.
\end{gather*}
The formula for the classical limit is analogous to the $so(4)$ case:
\[
 A=\lim_{\hbar\to0}e^{-\frac{\iu}{\hbar}(p_1x+p_2y)}\left(\hat{A}\,
 e^{\frac{\iu}{\hbar}(p_1x+p_2y)}\right),
\]
however, notice that here the Casimir operators (\ref{e3_la}) are pure
classical. Applying this procedure to the generators $M_i$ yields the bracket
(\ref{pq-e3}). In the Kowalevskaya case this procedure results in changing of a
coefficient in operator $\hat K$ (\ref{K.hatK}) (cf \cite{Laporte_1933}):
\[
 \hat K=\frac{1}{2}(\hat{k}_+\hat{k}_-+\hat{k}_-\hat{k}_+)
  +4\hbar^2(\hat{M}_1^2+\hat{M}_2^2).
\]

\subsection{The Clebsch top}\label{s:C}

There is no problem of ordering in this case and the quantum top is defined by
the Hamiltonians
\begin{align}
\label{C.Hq}
 \hat H&= \frac{1}{2}\sum_{i=1}^3\left(\hat{M}_i^2+\lambda_i\hat{\gamma}_i^2\right),\\
\label{C.Kq}
 \hat K&= \frac{1}{2}\sum_{i=1}^3
  \left(\lambda_i\hat{M}_i^2-\lambda\frac{\hat{\gamma}_i^2}{\lambda_i}\right),
  \quad \lambda=\lambda_1\lambda_2\lambda_3.
\end{align}
The use of representation (\ref{hat-e3}) yields the following commuting
differential operators:
\begin{align*}
 2\hat H_D&= -(x-y)^2D_xD_y+\frac{l}{a}(x-y)(D_x+D_y)\\
  &\quad +\frac{a^2z}{(x-y)^2}+a^2(\lambda_1+\lambda_2+\lambda_3),\\
 8\hat K_D&= r(x)D^2_x+2zD_xD_y+r(y)D^2_y\\
  &\quad +\Bigl(\frac{a-l}{2a}r'(x)+\frac{l}{a}z_y\Bigr)D_x
          +\Bigl(\frac{a+l}{2a}r'(y)-\frac{l}{a}z_x\Bigr)D_y\\
  &\quad +(\lambda_1-\lambda_2)\frac{l^2}{a^2}(x-y)^2
    -(\lambda_1-\lambda_2)\frac{l}{a}(x^2-y^2)+a^2\frac{r(x)r(y)-z^2}{(x-y)^4}
\end{align*}
where
\[
 r(x)=R(\lambda_1,\lambda_2,\lambda_3;x),\quad
 z=z(x,y)=W(\lambda_1,\lambda_2,\lambda_3;x,y).
\]
Notice, that in this case identity (\ref{dis}) takes the form
\begin{multline*}
z^2-r(x)r(y)=4(x-y)^4(\lambda_1\lambda_2+\lambda_2\lambda_3+\lambda_3\lambda_1)\\
 +4(x-y)^2W(\lambda_2\lambda_3,\lambda_3\lambda_1,\lambda_1\lambda_2;x,y),\qquad
\end{multline*}
therefore the last term in $\hat K_D$ partially cancels. The form of classical
Hamiltonians in the Darboux coordinates is analogous, with slightly different
coefficients:
\begin{align*}
 2H_D&= -(x-y)^2p_1p_2+\frac{2l}{a}(x-y)(p_1+p_2)\\
  &\quad +\frac{a^2z}{(x-y)^2}+a^2(\lambda_1+\lambda_2+\lambda_3),\\
 8K_D&= r(x)p^2_1+2zp_1p_2+r(y)p^2_2\\
  &\quad -\frac{l}{a}(r'(x)-2z_y)p_1+\frac{l}{a}(r'(y)-2z_x)p_2\\
  &\quad +4(\lambda_1-\lambda_2)\frac{l^2}{a^2}(x-y)^2+a^2\frac{r(x)r(y)-z^2}{(x-y)^4}.
\end{align*}

\subsection{Kowalevskaya top}\label{s:K}

The classical top is defined by Hamiltonians
\begin{equation}\label{K.HK}
 H=\frac{1}{2}(M_1^2+M_2^2+2M_3^2)-\frac{1}{2}\gamma_1,\quad K=k_+k_-
\end{equation}
where
\[
 k_{\pm}=(M_1\pm\iu M_2)^2+\gamma_1\pm\iu\gamma_2.
\]
The expressions in the Darboux coordinates (\ref{pq-e3}) are:
\begin{gather*}
 2H= x^2p_1^2+(4xy-x^2-y^2)p_1p_2+y^2p_2^2
    +\frac{2l}{a}(x-y)(p_1+p_2)+a\frac{xy-1}{x-y},\\
  K= \left(\left(x^2p_1+y^2p_2-\frac{2l}{a}(x-y)\right)^2-\frac{2axy}{x-y}\right)
     \left((p_1+p_2)^2+\frac{2a}{x-y}\right).
\end{gather*}
The quantum version of the Kowalevskaya top is of the form \cite{Laporte_1933,
Komarov_Kuznetsov_1987}
\begin{align}
\label{K.hatH}
 \hat H&= \frac{1}{2}(\hat M_1^2+\hat M_2^2+2\hat M_3^2)-\frac{1}{2}\hat\gamma_1,\\
\label{K.hatK}
 \hat K&= \frac{1}{2}(\hat{k}_+\hat{k}_-+\hat{k}_-\hat{k}_+)+4(\hat M_1^2+\hat M_2^2)
\end{align}
where
\[
 \hat k_{\pm}=(\hat M_1\pm\iu\hat M_2)^2+\hat\gamma_1\pm\iu\hat\gamma_2.
\]
The use of representation (\ref{hat-e3}) yields the following commuting
differential operators:
\begin{align*}
 2\hat H_D&= x^2D^2_x+(4xy-x^2-y^2)D_xD_y+y^2D^2_y,\\
 &\quad+\frac{1}{a}((a+l)x-ly)D_x+\frac{1}{a}(lx+(a-l)y)D_y+a\frac{xy-1}{x-y},\\
 \hat K_D&= \bigl[f^2-\frac{2axy}{x-y},g^2+\frac{2a}{x-y}\bigr]^+
   +4[f,gfg]^+ -2[fg,gf]^+ -2[f^2,g^2]^+
\end{align*}
where $[a,b]^+=\frac12(ab+ba)$ and
\[
 f=x^2D_x+y^2D_y-\frac{l}{a}(x-y),\quad g=D_x+D_y.
\]
It is worth noticing that Hamiltonians (\ref{K.hatH}), (\ref{K.hatK}) admit the
following generalization (Kowalevskaya gyrostat) \cite{Komarov_2001}:
\begin{align*}
 &2\hat H= \hat M_1^2+\hat M_2^2+2\hat M_3^2-\hat\gamma_1+c\hat M_3,\\
 &\begin{aligned}[b]
 \hat K&= \frac{1}{2}(\hat{k}_+\hat{k}_-+\hat{k}_-\hat{k}_+)
  +4(\hat M_1^2+\hat M_2^2)-2c(\hat M_1^2+\hat M_2^2)\hat M_3\\
 &\qquad +2c^2\hat M_3^2+c(c^2+1)\hat M_3-2c\hat M_1\hat\gamma_3
         -c^2\hat\gamma_1-\iu c\hat\gamma_2.
\end{aligned}
\end{align*}

\subsection{Goryachev--Chaplygin case}

The quantization was considered in \cite{Komarov_1982}. The Hamiltonians
\begin{align*}
 \hat H&= \hat M_1^2+\hat M_2^2+4\hat M_3^2-\hat\gamma_1+c\hat M_3,\\
 \hat K&= 4(\hat M^2_1+\hat M^2_2)\hat M_3+2\hat M_1\hat\gamma_3
     -4c\hat M^2_3+(1-c^2)\hat M_3+c\hat\gamma_1+\iu\hat\gamma_2
\end{align*}
satisfy the relation
\[
 [\hat H,\hat K]=4\iu l\hat M_2
\]
where $l=(\hat\gamma,\hat M)$ is one of the Casimir operators on $e(3)$.
Therefore, an integrable case occurs at $l=0$. The passage to operator
representation (\ref{hat-e3}) (at $l=0$) yields the commuting pair
\begin{gather*}
 \hat H_D = 3x^2D^2_x-(x^2-8xy+y^2)D_xD_y+3y^2D^2_y\\
  +(c+3)(xD_x+yD_y)+a\frac{xy-1}{x-y},\\
 -\hat K_D-4(\tfrac{c}{3}+1)\hat H_D =
  4x^3D^3_x+4(x^2+xy+y^2)(xD_x+yD_y)D_xD_y+4y^3D^3_y\\
   +(\tfrac{c}{3}+3)(4(x-y)^2D_xD_y-(c+3)(xD_x+yD_y))\\
   +\frac{a}{x-y}\Bigl((x+y)((x^2-1)D_x+(y^2-1)D_y)
  -(\tfrac{c}{3}+5)xy+\tfrac{c}{3}+3\Bigr).
\end{gather*}

\section{Spectra}

The quantization in terms of the generators of Lie algebra is universal, but
the setting of a boundary value problem and computation of the spectra depend
on the choice of a concrete representation. As an application, we consider here
the eigenvalue problem for the Euler top on $so(3)$ using representation
(\ref{hatUx}). Recall, that quantization of this model was obtained by
Kramers--Ittmann \cite{Kramers_Ittmann}. In the case of representation
(\ref{hatUx}), it is natural to define the spectrum by condition that
eigenfunctions are polynomial. This can be compared with Komarov--Kuznetsov
paper \cite{Komarov_Kuznetsov_1987} where the spectrum was found for the matrix
representation, and the recent Grosset--Veselov paper \cite{Grosset_Veselov}
where the spectrum was studied for the representation in elliptic coordinates
and it was shown that coefficients of the characteristic polynomial at a given
level set $j=s$ are expressed through the so-called elliptic Bernoulli
polynomials. The spectral problem for the tops on $so(4)$ is rather complicated
and we restrict ourselves by derivation of equations for eigenfunctions in the
Schottky--Manakov case.

\subsection{Matrix representation of $so(3)$}

Let us introduce the wave function $|m,j\rangle=x^{j-m}$, then representation
(\ref{hatUx}) has the following matrix elements in this basis:
\begin{equation}\label{so3:our}
\begin{gathered}[b]
 \hat U_1\pm\iu\hat U_2=\hat U_\pm,\quad \hat{U}_3|m,j\rangle=m|m,j\rangle,\\
 \hat{U}_+|m,j\rangle=(j-m)|m+1,j\rangle,\quad
 \hat{U}_-|m,j\rangle=(j+m)|m-1,j\rangle.
\end{gathered}
\end{equation}
In this case the condition that $j-m$ is integer follows from the condition
that the basis functions $x^{j-m}$ must be single-valued. If the problem admits
the time-reversal symmetry $m\to-m$ then $j+m$ should be integer as well, and
this implies that $j$ and $m$ are simultaneously integer or half-integer. The
form of the matrix elements implies that $m=-j,\dots,j$, and the condition
$j>0$ follows from the condition that the basis function must be analytic. It
is clear from its form that wave function on the orbit $j=\const$ is a
polynomial of degree $2j$.

The wave function and the spectral problem on $so(3)$ can be written as
follows:
\[
 |\psi\rangle=\sum\limits_{m=-j}^{j}C(m,j)\,|m,j\rangle,\quad
 H\,|\psi\rangle=\lambda\,|\psi\rangle
\]
Analogously, the wave function and the spectral problem on $so(4)$ read:
\[
 |\psi\rangle=\sum_{m_1=-j_1}^{j_1}
 \sum_{m_1=-j_2}^{j_2}C(m_1,m_2;j_1,j_2)\,|m_1,m_2,j_1,j_2\rangle,\quad
 H\,|\psi\rangle=\lambda\,|\psi\rangle.
\]
In the differential representation, the basis wave function on $so(4)$ is
chosen as $|m_1,m_2,j_1,j_2\rangle=x^{j_1-m_1}y^{j_2-m_2}$. Then the general
wave function is a polynomial of degree $2j_1$ with respect to $x$ and degree
$2j_2$ with respect to $y$.

\subsection{Spectrum of the Euler top on $so(3)$}

The eigenvalue problem is of the form
\[
 \hat{H}\psi_j^{\lambda}(x)=\lambda\psi_j^{\lambda}(x),\quad
 \hat{C}\psi_j^{\lambda}(x)=j(j+1)\psi_j^{\lambda}(x)
\]
where
\[
 \hat{H}=\alpha_1\hat{U}_1^2+\alpha_2\hat{U}_2^2+\alpha_3\hat{U}_3^2,\quad
 \hat{C}=\hat{U}_1^2+\hat{U}_2^2+\hat{U}_3^2.
\]
We will assume that $\alpha_1+\alpha_2+\alpha_3=0$, up to an unessential shift
of the spectrum.

Let us use the differential operators representation for $so(3)$ algebra
(\ref{so3:our}) and represent wave function in the form
\[
\psi_j^{\lambda}(x)=\sum\limits_{k=0}^{2j}\tilde C_j^{\lambda}(k)x^k.
\]
Then the eigenvalue problem is rewritten in the form of recurrent relation
\begin{gather*}
 \frac{1}{4}(2j+1-k)(2j+2-k)C_j^{\Lambda}(k-2)\\
 +\Bigl(\frac{1}{2}(j(j+1)-3(j-k)^2)\xi
 -\Lambda\Bigr)C_j^{\Lambda}(k)\\
 +\frac{1}{4}(k+1)(k+2)C_j^{\Lambda}(k+2)=0
\end{gather*}
where $\xi=\frac{\alpha_1+\alpha_2}{\alpha_1-\alpha_2}$,
$\Lambda=\frac{\lambda}{\alpha_1-\alpha_2}$, $C_j^{\Lambda}=\tilde
C_j^{\lambda}$, with the boundary conditions on the left end
\[
 C_j^{\Lambda}(-2)=C_j^{\Lambda}(-1)=0,\quad
 C_j^{\Lambda}(0)=C_j^{\Lambda}(1)=1.
\]
The problem is splitting for odd and even polynomials.

If $j$ is integer then the boundary conditions on the right end is
$C_j^{\Lambda}(2j+2)=0$ for the even polynomials and $C_j^{\Lambda}(2j+1)=0$
for the odd ones.

If $j$ is half-integer then, vice-versa, the boundary conditions on the right
end is $C_j^{\Lambda}(2j+1)=0$ for the even polynomials and
$C_j^{\Lambda}(2j+2)=0$ for the odd ones.

As a result, the eigenvalues of the Hamiltonian $\hat{H}$ are zeroes of the
polynomial
\[
 P_j(\Lambda)=C_j^{\Lambda}(2j+1)C_j^{\Lambda}(2j+2),\quad
 \deg P_j(\Lambda)=2j+1.
\]
Let us explicitly write down several polynomials $P_j(\Lambda)$ normalized by
the condition that the coefficient of the leading term $\Lambda^{2j+1}$ is
unit:
\begin{align*}
 & P_0(\Lambda)=\Lambda,\\
 & P_{1/2}(\Lambda)=\Lambda^2,\\
 & P_1(\Lambda)=\frac{1}{4}(\Lambda-\xi)(2\Lambda+\xi+1)(2\Lambda+\xi-1),\\
 & P_{3/2}(\Lambda)=\frac{1}{16}(4\Lambda-9\xi^2-3)^2,\\
 & P_2(\Lambda)=\frac{1}{4}(\Lambda+3\xi)(2\Lambda-3\xi+3)(2\Lambda-3\xi-3)(\Lambda-9\xi^2-3)^2,\\
 & P_{5/2}(\Lambda)=(\Lambda^3-7\Lambda(3\xi^2+1)+20\xi(\xi^2-1))^2.
\end{align*}
All polynomials $P_j(\lambda)$ with half-integer $j$ are full squares in virtue
of the Kramers theorem \cite[p. 225]{LL} about the double degeneration of the
systems with half-integer value of the spin.

\subsection{Matrix representation for the Schottky--Manakov top}

In the Schottky--Manakov case we have two consistent eigenvalue problems
$\hat{H}\psi=\lambda\psi$, $\hat{K}\psi=\mu\psi$. The wave function is of the
form
\[
 \psi_{j_1,j_2}^{\lambda,\mu}(x,y)=
 \sum_{k=0}^{2j_1}\sum_{l=0}^{2j_2}C_{j_1,j_2}^{\lambda,\mu}(k,l)x^ky^l.
\]
In the lattice representation, the 5-point equation appears for the eigenvalues
$\mu$:
\begin{align*}
 &~~~~ (\alpha_1-\alpha_2)(2j_1+1-k)(2j_2+1-l)C_{j_1,j_2}^{\lambda,\mu}(k-1,l-1)\\
 &+(\alpha_1-\alpha_2)(k+1)(l+1)C_{j_1,j_2}^{\lambda,\mu}(k+1,l+1)\\
 &+2(2(j_1-k)(j_2-l)\alpha_3-\mu)C_{j_1,j_2}^{\lambda,\mu}(k,l)\\
 &+(\alpha_1+\alpha_2)(2j_1+1-k)(l+1)C_{j_1,j_2}^{\lambda,\mu}(k-1,l+1)\\
 &+(\alpha_1+\alpha_2)(k+1)(2j_2+1-l)C_{j_1,j_2}^{\lambda,\mu}(k+1,l-1)=0,
\end{align*}
and 9-point one for the eigenvalues $\lambda$:
\begin{align*}
 &~~~~ (\alpha_2^2-\alpha_1^2)(k+1)(k+2)C_{j_1,j_2}^{\lambda,\mu}(k+2,l)\\
 &+(\alpha_2^2-\alpha_1^2)(l+1)(l+2)C_{j_1,j_2}^{\lambda,\mu}(k,l+2)\\
 &+2\alpha_3(\alpha_2-\alpha_1)(k+1)(l+1)C_{j_1,j_2}^{\lambda,\mu}(k+1,l+1)\\
 &+2\alpha_3(\alpha_2+\alpha_1)(k+1)(2j_2+1-l)C_{j_1,j_2}^{\lambda,\mu}(k+1,l-1)\\
 &+\Bigl((\alpha_1-\alpha_2)^2\bigl((2j_1+1-k)^2+(2j_2+1-l)^2\\
 &\qquad -2(j_1+j_2+1-k)(j_1+j_2+1-l)\\
 &\qquad -(j_1-j_2)^2-j_1(j_1+1)-j_2(j_2+1)\bigr)\\
 &\qquad -4\alpha_3^2((j_1-k)^2+(j_2-l)^2)\\
 &\qquad +(\alpha_2+\alpha_1)^2((j_1+j_2-k-l)^2-j_1(j_1+1)-j_2(j_2+1))\Bigr)
  C_{j_1,j_2}^{\lambda,\mu}(k,l)\\
 &+(\alpha_2^2-\alpha_1^2)(2j_1+1-k)(2j_1+2-k)C_{j_1,j_2}^{\lambda,\mu}(k-2,l)\\
 &+(\alpha_2^2-\alpha_1^2)(2j_2+1-l)(2j_2+2-l)C_{j_1,j_2}^{\lambda,\mu}(k,l-2)\\
 &+2\alpha_3(\alpha_2+\alpha_1)(2j_1+1-k)(l+1)C_{j_1,j_2}^{\lambda,\mu}(k-1,l+1)\\
 &+2\alpha_3(\alpha_2-\alpha_1)(2j_1+1-k)(2j_2+1-l)C_{j_1,j_2}^{\lambda,\mu}(k-1,l-1)=0.
\end{align*}
The boundary conditions are: $C_{j_1,j_2}^{\lambda,\mu}(k,l)=0$ if the pair
$(k,l)$ lies outside the rectangle with the vertices $(0,0)$, $(j_1,0)$,
$(j_1,j_2)$, $(0,j_2)$.

Solutions $C_{j_1,j_2}^{\lambda,\mu}(k,l)$ split into solutions on two
sublattices: a solution is called ``even'' if it vanishes at odd $k+l$, and it
is called ``odd'' if it vanishes at even $k+l$.

Actually, one can avoid solving 9-point equation: it is sufficient to determine
the wave functions from the 5-point equation and then the substitution into the
9-point one allows to determine the relation between $\lambda_i$ and $\mu_i$. A
plausible answer is that the pairs $(\lambda_i,\mu_i)$ lie on a certain
algebraic curve.

Consider the case $j_1=j$, $j_2=\frac{1}{2}$ as an example. The wave function
is a superposition of odd and even ones. The even wave function is of the form
\[
 \psi=C_0+C_1xy+C_2x^2+C_3x^3y+\dots
\]
and coefficients satisfy the boundary conditions $C_{-2}=C_{-1}=C_{2j+1}=0$ and
recurrent relations
\begin{multline*}
 \frac{1}{2}(\alpha_1+(-1)^k\alpha_2)(2j+1-k)C_{k-1}+(\alpha_3(j-k)(-1)^k-\mu)C_k \\
 +\frac{1}{2}(\alpha_1-(-1)^k\alpha_2)(k+1)C_{k+1}=0.
\end{multline*}
Up to the constant factors, one finds
\[
 C_0=0,\quad C_1=\mu-j\alpha_3,\quad
 C_2=\mu^2-\alpha_3\mu-\alpha_3^2j(j-1)-\frac{1}{2}j(\alpha_1-\alpha_2)^2,\quad \dots
\]
The odd wave function is of the form
\[
\psi=B_0y+B_1x+B_2x^2y+B_3x^3+\dots,
\]
the coefficients satisfy the boundary conditions $B_{-2}=B_{-1}=B_{2j+1}=0$ and
recurrent relations
\begin{multline*}
 (\alpha_1-(-1)^k\alpha_2)(2j+1-k)B_{k-1}+2(\alpha_3(k-j)(-1)^k-\mu)B_k\\
 +(\alpha_1+(-1)^k\alpha_2)(k+1)B_{k+1}=0.
\end{multline*}
Up to the constant factors, one finds
\[
 B_0=0,\quad B_1=\mu+j\alpha_3,\quad
 B_2=\mu^2+\alpha_3\mu-\alpha_3^2j(j-1)-\frac{1}{2}j(\alpha_1-\alpha_2)^2,\quad\dots
\]
The eigenvalues are found from equation
\[
 P_{j,1/2}(\mu)=C(2j+1)B(2j+1)=0,\quad \deg P_{j,1/2}(\mu)=2(2j+1).
\]
The polynomial $P_{j,1/2}(\mu)$ is a full square if $j$ is integer, in
accordance with the Kramers theorem, because then $j+\frac{1}{2}$ is
half-integer.

\section*{Acknowledgements}
\addcontentsline{toc}{section}{Acknowledgements}

The authors thank I.M. Krichever and V.V. Sokolov for their interest to this
work and useful remarks. The research was supported by grants NSh--6501.2010.2
and RFBR 10-01-00088.


\end{document}